\documentclass[preprint,aps,showpacs]{revtex4}
\usepackage{epsfig}
\usepackage{psfrag}

\newcommand{\ket}[1]{| #1 \rangle}
\newcommand{\bra}[1]{\langle #1 |}

\newcommand{\br}{{\bf r}}

\newcommand{\bK}{{\bf K}}

\newcommand{\bk}{{\bf k}}
\newcommand{\bq}{{\bf q}}
\newcommand{\bd}[1]{{\bf{#1}}}
\newcommand{\brho}{{\bf \rho}}

\psfrag{figure1ylabel}{$g^\prime$ (arb. units)}
\psfrag{figure1xlabel}{$l_z/a_{\rm 3d}$}
\psfrag{figure2aylabel}{$\psi(\rho)$}
\psfrag{figure2axlabel}{$\rho$ (units of $l_\rho$)}
\psfrag{figure2bylabel}{$\psi(\rho)$}
\psfrag{figure2bxlabel}{$\rho$ (units of $l_\rho$)}
\psfrag{figure3ylabel}{$10^{-3}N_0$}
\psfrag{figure3xlabel}{$\mu$ (units of $\hbar \omega$)}
\psfrag{figure4ylabel}{$\phi(\rho)$}
\psfrag{figure4xlabel}{$\rho$ (units of $l_\rho$)}
\psfrag{figure5ylabel}{$\Omega_c$ (units of $\omega$)}
\psfrag{figure5xlabel}{$10^{-3}N_0$}

\begin{document}

\title{Energy dependent scattering and the Gross-Pitaevskii Equation in two
dimensional Bose-Einstein condensates}
\author{M.D.~Lee, S.A.~Morgan, M.J.~Davis, and K.~Burnett} 
\affiliation{Clarendon Laboratory, Department of Physics, University of Oxford,
Parks Road, Oxford OX1 3PU, United Kingdom} 
\date{\today}

\begin{abstract} 
We consider many-body effects on particle scattering in one, two and three
dimensional Bose gases.  We show that at $T=0$ these effects can be modelled by
the simpler two-body T-matrix evaluated off the energy shell.  This is
important in 1D and 2D because the two-body T-matrix vanishes at zero energy
and so mean-field effects on particle energies must be taken into account to
obtain a self-consistent treatment of low energy collisions.  Using the
off-shell two-body T-matrix we obtain the energy and density dependence of the
effective interaction in 1D and 2D and the appropriate Gross-Pitaevskii
equations for these dimensions.  Our results provide an alternative derivation
of those of Kolomeisky \emph{et al.\ }\cite{kolomeisky2000,kolomeisky1992}.  We
present numerical solutions of the Gross-Pitaevskii equation for a 2D
condensate of hard-sphere bosons in a trap.  We find that the interaction
strength is much greater in 2D than for a 3D gas with the same hard-sphere
radius. The Thomas-Fermi regime is therefore approached at lower condensate
populations and the energy required to create vortices is lowered compared to
the 3D case.
\end{abstract}

\maketitle

\section{Introduction}

Recent experiments on the quasicondensation of a two dimensional gas of atomic
hydrogen~\cite{safonov1998} and the possibilities of confining dilute atomic
gases in ``low-dimensional'' traps~\cite{gorlitz2001,lowdimensional} have
stimulated interest in the possibilities of Bose-Einstein condensation in
two-dimensional systems.  It has long been known that, in the thermodynamic
limit, Bose condensation is not possible in two dimensional homogeneous systems
at any finite temperature because long wavelength fluctuations destroy long range
coherence~\cite{hohenberg1967}.  Instead such a system undergoes a
Kosterlitz-Thouless transition~\cite{Kosterlitz1973} and acquires local
coherence properties over a length scale dependent on the temperature --- a
``quasicondensate''~\cite{popov}.  In the limit $T \rightarrow 0$ global
coherence is achieved in homogeneous 2D systems and a true condensate then
exists. In a trapped 2D system the modifications of the density of states
caused by the confining potential enable a true condensate to exist even at
finite temperatures~\cite{bagnato1991}.

In most treatments of a Bose condensed gas in 3D, particle interactions are
described by a delta-function contact potential whose strength is determined by
the zero energy and momentum limit of the two-body T-matrix ($T_{\rm 2b}$)
which describes scattering in a vacuum.   This leads to the standard form of
the interaction potential $\left(4\pi\hbar^2 a_{\rm 3d}/m \right)\delta(\br)$,
where $a_{\rm 3d}$ is the \emph{s}-wave scattering length.  At higher order it
can be shown that the interactions are actually described by a many-body
T-matrix ($T_{\rm MB}$)~\cite{morgan2000,stoof1997,shi1998} which accounts for
the fact that collisions occur in the presence of the condensate rather than in
free space.  In 2D this correction is critical because the 2D two-body T-matrix
vanishes in the zero energy limit~\cite{fisher,kolomeisky1992}, and thus we
must include this correction (at least partially) even at leading
order~\cite{schick}. In this paper we develop an expression for the many-body
T-matrix in terms of the two-body T-matrix evaluated at a shifted effective
interaction energy. In one and two dimensions we obtain an effective
interaction which depends on the energy of the collision, in contrast with
three dimensional gases.   

The energy dependence of the effective interaction can be written as a density
dependence, in which form the results can be applied to trapped gases.  This
leads to a Gross-Pitaevskii equation (GPE) describing the condensate wave
function which no longer has a cubic non-linearity in $\psi$, but instead goes
as $|\psi|^4\psi$ in 1D and as $(|\psi|^2/\ln|\psi|^2)\psi$ in 2D.  Such a
modified GPE has already been introduced by
Kolomeisky~\cite{kolomeisky2000,kolomeisky1992} and Tanatar~\cite{tanatar2000},
using arguments based either on the renormalisation group or a Kohn-Sham
density functional approach~\cite{nunes1999}.  Our discussion in this paper is
to show how essentially the same results can be obtained by a consideration of
many-body effects on particle scattering and to relate this to well-understood
treatments of the 3D Bose gas.  Indeed, substantially the same treatment as
used in 3D applied to the 1D and 2D gases leads to the energy dependent
effective interactions.  The principle difference is that these effects must be
taken into account in leading order, whereas in 3D they can be neglected in the
simplest treatments and only become important at finite temperature or high
density.

In the following section we discuss the Gross-Pitaevskii equation, and the
limits in which a system may be considered two dimensional.  In
section~\ref{sec:Tmatrix} we then derive the many-body effective interaction
for low dimensional gases, before considering its implications for 1D gases in
section~\ref{sec:1D}.  Finally, using this effective interaction we obtain a
form of the two dimensional Gross-Pitaevskii equation, and we present the
results of numerical solutions for both ground and vortex states in
section~\ref{sec:2Dnumerical}.

\section{The Gross-Pitaevskii Equation in 2D and quasi-2D}

\label{sec:quasi2d}

The macroscopic wave function for a Bose-Einstein condensate is found in mean
field theory using a non-linear Schr\"{o}dinger equation known as the
Gross-Pitaevskii equation --- where the non-linear term arises from
interactions between the atoms of the condensate.  Obtaining the form of the
effective interaction in 2D, and describing its effect on the solutions of this
equation are the main concerns of this paper.

Currently, most BEC experiments have created three
dimensional condensates, which are described by a GPE of the form
\begin{equation}
-{\hbar^2 \over 2m} \nabla^2 \psi(\br) + V_{\rm trap}(\br) \psi(\br) +
N_0 
g_{\rm 3d}\left|
\psi(\br)\right|^2 \psi(\br) = \mu \psi(\br), \label{eq:3DGPE}
\end{equation}
where $V_{\rm trap}(\br)$ is the external trapping potential, $N_0$ is the
condensate population, $\mu$ the chemical potential, and $g_{\rm 3d}$ is the
coupling parameter describing the effective interactions.  The coupling
parameter is generally taken to be the zero energy and momentum limit of the
two body T-matrix which in 3D is a non-zero constant $g_{\rm 3d} = 4\pi\hbar^2
a_{\rm 3d}/m$, where $a$ is the \emph{s}-wave scattering length. The T-matrix
has the contact potential form $T_{\rm 2b}(\br,\br') = g_{\rm 3d}
\delta(\br-\br')\delta(\br)$ in the limit that all the momenta involved in
typical collisions are much less than $1/R_e$, where $R_e$ is the range of the
actual interatomic potential (which is not in general equal to the scattering
length $a$).

The obvious extension of these experiments in order to achieve the goal of two
dimensional condensates is to confine a gas in an anisotropic trap such that
the gas is tightly confined in the $z$ direction. For a harmonic potential such
a trap has the form $V_{\rm trap}(\br) = m\omega^2/2\left({\bf
\rho}^2+z^2/\gamma\right)$, with $l_z \equiv \sqrt{\hbar/2m\omega_z}$ as the
characteristic trap length in the tightly confined direction, where $\omega_z
\equiv \omega/\gamma^{1/2}$.  On decreasing $l_z$ (decreasing $\gamma$) the
system will pass from being three dimensional to being two dimensional in a
variety of senses.

The system can first be called two dimensional once $l_z$ has merely been
decreased sufficiently that the mean-field energy of the condensate is small
compared to $\hbar\omega_z$.  In this case the dynamics of the system in the
$z$ dimension are restricted to zero point oscillations.  Nonetheless, if $l_z$
is still much greater than $a_{\rm 3d}$, then two body collisions are hardly
affected, and hence interactions can still be described by the three
dimensional contact potential $g_{\rm 3d}$.   Therefore, although in this case
the third dimension can be factored out of the dynamics of the system, at short
length scales the interactions are still three dimensional.  This regime can be
described using the 3D GPE of Eq.~(\ref{eq:3DGPE}) with the assumption that the
wave function can be factorised as 
\begin{equation}
\Psi({\bf \rho},z) = \psi({\bf \rho})\left({m \omega_z  \over
\pi\hbar}\right)^{1/4}\exp\left(-{m\omega_z\over 2\hbar}{z}^2\right).
\end{equation} 
Substituting into the 3D GPE, and integrating over $z$ leads to
a two dimensional equation 
\begin{equation} -{\hbar^2 \over 2m}\nabla_{\bf
\rho}^2\psi({\bf \rho})+  {1\over 2} m \omega^2 {\bf \rho}^2\psi({\bf \rho}) +
g'N_0 \left| \psi({\bf \rho}) \right|^2\psi({\bf \rho})  = \mu' \psi({\bf
\rho}), \label{eq:q2dGPE} 
\end{equation} 
where $\rho = \{x,y\}$, $\mu' = \mu -
\hbar\omega_z/2$, and the coupling parameter $g'$ is given by $g'_{\rm 3d} =
(m\omega_z/2\pi\hbar)^{1/2}g_{\rm 3d}$.  The subscript here refers to the three
dimensional nature of the interactions whilst the prime indicates that $g'_{\rm
3d}$ is a two dimensional quantity. 

The above factorisation of the wave function remains valid as $l_z$ is
decreased further, but the assumption that the scattering is unaffected begins
to break down when $l_z$ is not much greater than $a_{\rm 3d}$.   The effect of
the confinement on particle interactions has been discussed in detail by Petrov
and Shlyapnikov~\cite{petrov2000,petrov2001}, who found that a 2D contact
potential can still be used but that the strength of the interaction becomes
dependent upon the confinement.  The coupling parameter which they obtained in
this ``quasi-2D'' regime is
\begin{equation}
g'_{\rm q2d} = \left({8\pi\omega_z\hbar^3 \over m}\right)^{1/2}
\left[{1\over a_{\rm 3d}} + \left({m \omega_z \over2\pi\hbar}\right)^{1/2}
\ln\left({B\hbar\omega_z\over2\mu\pi}\right)\right]^{-1}, \label{eq:gq2d}
\end{equation}
where $B \approx 0.915$.  This expression is valid when the conditions
$mg'_{\rm q2d}/2\pi\hbar^2, R_e/l_z, 2\mu/\hbar\omega_z \ll 1$ are satisfied. 
In the large $l_z$ limit the $1/a_{\rm 3d}$ term dominates and the scattering
is three dimensional as considered above.  However in the fully 2D limit the
logarithmic term in Eq.~(\ref{eq:gq2d}) dominates and $g$ becomes
dependent upon $\mu$.  Equation~(\ref{eq:gq2d}) was derived from solving the
two body scattering problem within the potential causing the tight $z$
confinement.  We will now show how essentially the same result can be obtained
in the fully 2D limit by a consideration of the many-body effects on particle
scattering.

\section{The T-matrix in the GPE}
\label{sec:Tmatrix}

In order to describe the interactions within a truly 2D BEC we must consider 2D
scattering in the presence of a condensate. This is described by a many body
T-matrix $T_{\rm MB}$, and the coupling parameter which appears in the GPE is
in fact given by the matrix element $\bra{\bk'}T_{\rm MB}(E)\ket{\bk}$
evaluated in the limit of zero momentum and energy ($\bk,\bk',\bd{K},E =0$). 
Note that the many-body T-matrix is in principle also a function of the centre
of mass momentum $\bd{K}$, but this will not be explicitly indicated in this
paper for notational simplicity.  This will not be important for the results
presented since we will always take the limit $\bd{K} = \bd{0}$ in this paper.

Before discussing the many-body T-matrix, however, we will first consider the
simpler two-body T-matrix which describes collisions between two particles in a
vacuum and for which analytical expressions exist~\cite{scattering} .  We will
then show how  the many-body T-matrix can be obtained from the
two-body version in the limit appropriate for the study of BEC.

\subsection{The two-body T-matrix}

The two-body T-matrix describing scattering from an interparticle potential
$V(\br)$ is the solution to the Lippmann-Schwinger
equation~\cite{taylor}
\begin{eqnarray}
\bra{\bk'}T_{\rm 2b}(\bar{E})\ket{\bk} &=& 
\bra{\bk'}V(\br_1-\br_2)\ket{\bk} \nonumber \\
&&+
\sum_\bq
\bra{\bk'}V(\br_1-\br_2)\ket{\bq}{1 \over \bar{E} - 
(\varepsilon^{sp}_{\bK/2+\bq} +
\varepsilon^{sp}_{\bK/2-\bq})}
\bra{\bq}T_{\rm 2b}(\bar{E})\ket{\bk}, \label{eq:2bTLS}
\end{eqnarray}
where $\bk$ and $\bk'$ are the relative momenta of the two particles before and
after the collision respectively, and $\bd{K}$ is the centre-of-mass momentum. 
The energy of a single particle state is $\varepsilon^{sp}_k$, where in the
homogeneous limit  $\varepsilon_k^{sp} = \hbar^2k^2/2m$.  The total energy of
the collision is $\bar{E}$ and includes a contribution from the centre-of-mass
momentum $\bd{K}$ which cancels the corresponding contribution from the single
particle energies.  The two-body T-matrix is therefore independent of $\bd{K}$,
as it must be in free space.

The scattering event described here could be a single interaction
$\bra{\bk'}V\ket{\bk}$, or alternatively the particles may first make a
transition to an intermediate state $\ket{\bq}$ (weighted by an energy
dependent denominator) before interacting again to emerge in state
$\ket{\bk'}$.  The recursive nature of Eq.~(\ref{eq:2bTLS}) sums all
possible processes for which $\ket{\bk} \rightarrow \ket{\bk'}$.  For many
applications we only need the ``on-shell'' T-matrix where both the energy and
momentum conservation laws are fulfilled. However, it is also useful to
consider the more general off-shell form shown above, where the momenta and
energy may take arbitrary values.

It can be shown that, for interaction potentials of a finite range $R_e$, the
T-matrix is independent of the incoming and outgoing momenta (in the limit
$kR_e,k'R_e \ll 1$)~\cite{scattering}. In the position representation this
corresponds to an effective interaction which is proportional to
$\delta(\br_1-\br_2)$.  This contact potential approximation is of great
utility in solving the GPE where the zero momentum limit of the T-matrix is
used to describe particle interactions.  In the three dimensional case the
T-matrix elements at low energy and momenta are also independent of energy,
leading to a constant coupling parameter in the GPE with form $g_{\rm 3d}
=4\pi\hbar^2a_{\rm 3d}/m$ in Eq.~(\ref{eq:3DGPE}).

The contact potential approximation is still valid in one and two dimensions,
but the T-matrix at leading order now depends upon the energy of the collision,
as will be shown in the following sections. Thus the scattering terms in the 2D
GPE will be quite different from the three dimensional case.

\subsection{The many-body T-matrix}

The two-body T-matrix describes collisions in vacuo in which the intermediate
states are single-particle in nature.  However, in a Bose condensed gas
collisions occur in the presence of a condensate and a many-body T-matrix is
needed to describe scattering processes.  This is defined by the
equation
\begin{eqnarray}
\bra{\bk'}T_{\rm MB}(E)\ket{\bk} &=& 
\bra{\bk'}V(\br_1-\br_2)\ket{\bk} \nonumber \\
&&+
\sum_\bq
\bra{\bk'}V(\br_1-\br_2)\ket{\bq}{(1 +n_{\bK/2+\bq} +n_{\bK/2-\bq})
 \over E - (\varepsilon_{\bK/2+\bq} +
\varepsilon_{\bK/2-\bq})}
\bra{\bq}T_{\rm MB}(E)\ket{\bk}, \label{eq:MBTLS}
\end{eqnarray}
where $E$ is the interaction energy, and $\varepsilon_{\bq}$ is the energy of a
quasiparticle state of momentum $\bq$, which is given by  
\begin{equation}
\varepsilon_p = \left[(\varepsilon_p^{\rm sp})^2 + 2\varepsilon_p^{\rm sp}\mu
\right]^{1/2}, \label{eq:bogenergies}
\end{equation}
in the Bogoliubov approximation~\cite{bogoliubov1947} for the case of the hard
sphere gas.  The corrections included in this many-body T-matrix over the
two-body version are  the occurrence of quasiparticle rather than particle
energies for the intermediate states, and the Bose enhancement of scattering
into these states.  This latter effect results in the presence of population
factors $n_{\bq}$ in Eq.~({\ref{eq:MBTLS}).

Formally, this many-body T-matrix is included in the theory of a Bose condensed
gas by considering the effect of the so-called anomalous average $\langle
\hat{a}_i\hat{a}_j \rangle$ on the condensate evolution, where $\hat{a}_i$ is
the non-condensate annihilation operator for state $i$.  This term occurs when
terms in the Hamiltonian of higher than quadratic order in
$\hat{a}_i,\hat{a}_i^\dagger$ are taken into account~\cite{morgan2000,GHFB}. 
We note that a generalisation of the many-body T-matrix which includes
quasiparticle propagator factors for the intermediate states has been
proposed~\cite{stoof1997}, but the corrections this includes over and above
Eq.~(\ref{eq:MBTLS}) are of still higher order.

We note that the energies $\varepsilon_{\bq}$ and $E$ in $T_{\rm MB}$ are
measured relative to the condensate, whereas the single particle energies in
$T_{\rm 2b}$ are measured relative to the energy of a stationary particle. This
means that for collisions between particles in the condensate we take the limit
$E=0$ in $T_{\rm MB}$, which corresponds to $\bar{E} = 2\mu$ when measured
relative to the same zero of energy as the two body
case~\cite{morgan2000}.  For collisions between condensate atoms, we also take
the zero momentum limit $\bk,\bk',\bd{K}=\bd{0}$.  Interactions between two
condensate atoms are therefore described by the matrix element
$\bra{\bd{0}}T_{\rm MB}(0)\ket{\bd{0}}$.

\subsection{$T_{\rm MB}$ in terms of $T_{\rm 2b}$, a simple argument}

\label{sec:negenergy}

The Lippmann-Schwinger equation for the many-body T-matrix is substantially
more difficult to solve than the two-body equivalent due to the presence of
quasiparticle energies and populations.   In the limit of zero temperature we
will show that the many body T-matrix can be approximated by an off shell two
body T-matrix evaluated at a negative energy.  To see this we consider
Eq.~(\ref{eq:MBTLS}) for the matrix element $\bra{\bd{0}}T_{\rm
MB}(0)\ket{\bd{0}}$ at $T=0$ where the population terms vanish.  Upon
comparison with Eq.~(\ref{eq:2bTLS}) it can be seen that the only
difference between the equations for the two types of T-matrix occurs in the
energy denominators.  Specifically, the quasiparticle energy spectrum appears
in the many-body case, whereas the single particle spectrum appears in the
two-body case.  Heuristically, if the dominant contribution to the intermediate
states in a collision comes from states with energies of order $\mu$ or higher,
we can proceed by replacing $\varepsilon_k$ by $\varepsilon^{sp}_k+\mu$. This
is the high energy limit of the Bogoliubov spectrum of
Eq.~(\ref{eq:bogenergies}) and it contains a constant shift from the
single particle spectrum due to the mean field effects of the condensate which
do not vanish in the relevant momentum range $k \sim k_0$ for a contact
potential interaction (where $\hbar^2k_0^2/2m \equiv \mu$).  We are interested
in the many-body T-matrix at $E=0$, and thus the energy denominator in
Eq.~(\ref{eq:MBTLS}) becomes
\begin{eqnarray}
{1 \over 0 - (\varepsilon_{\bK/2+\bq} + \varepsilon_{\bK/2-\bq})} &\approx &
{1 \over 0 - (\varepsilon^{sp}_{\bK/2+\bq} +\mu +
\varepsilon^{sp}_{\bK/2-\bq}+\mu )}  \nonumber \\
&=& {1 \over -2\mu - (\varepsilon^{sp}_{\bK/2+\bq} +
\varepsilon^{sp}_{\bK/2-\bq})}.
\end{eqnarray}
Comparison with Eq.~(\ref{eq:2bTLS}) for the two-body T-matrix shows that
in this approximation
\begin{equation}
\bra{\bd{0}}T_{\rm MB}(0)\ket{\bd{0}} = 
\bra{\bd{0}}T_{\rm 2b}(-2\mu)\ket{\bd{0}}. \label{eq:Estarsimple}
\end{equation}
Interestingly, this shows that the effective two body interaction energy is
negative, meaning that the interaction strength is always real.  We will see
that this is important in the 1D case in the following section.  In 3D the
two-body T-matrix is independent of energy to first order, but in both one and
two dimensions it has a non-trivial energy dependence and therefore the
effective interaction energy becomes important in these lower dimensions.

At first glance the result of Eq.~(\ref{eq:Estarsimple}) may appear
counterintuitive since the energy of a collision between two condensate
particles might be thought to be $+2\mu$, and certainly not negative.  However,
as we have shown, the many-body effects in the system lead to a shift in the
quasiparticle energy spectrum and it is this which leads to a  shifted
effective energy entering the two-body T-matrix.   Stoof \emph{et al.\
}\cite{stoof1993,anderson2001}  have also proposed that interactions in
low-dimensional condensates can be described by the two-body T-matrix evaluated
at a negative energy ($-2\mu$), the same result given by our heuristic argument
above.  In the following section we will use a more rigorous argument and find
that this leads to somewhat better values for the effective interaction energy.

\subsection{$T_{\rm MB}$ in terms of $T_{\rm 2b}$, a better argument}

Having shown heuristically in the previous section that the many-body T-matrix
can be approximated by a two-body T-matrix evaluated at a negative energy, we
will now present a more formal justification.  This will lead to a slight
modification to the magnitude of the energy used in the two body T-matrix,
but the essential physics of the argument is unchanged.

From
equations~(\ref{eq:2bTLS}) and~(\ref{eq:MBTLS}), it is possible to derive an
expression for the many-body T-matrix solely in terms of the two-body
T-matrix~\cite{morgan2000}
\begin{equation}
\bra{\bk'}T_{\rm MB}(E)\ket{\bk} = \bra{\bk'}T_{\rm 2b}(\bar{E})\ket{\bk} +
\bra{\bk'}T_{\rm corr}(E,\bar{E})\ket{\bk},
\end{equation}
where
\begin{eqnarray}
\bra{\bk'}T_{\rm corr}(E,\bar{E})\ket{\bk} &=&
\sum_{\bq \neq 0}{\bra{\bk'}T_{\rm 2b}(\bar{E})\ket{\bq}
(1 +n_{\bK/2+\bq} +n_{\bK/2-\bq})\bra{\bq}T_{\rm MB}(E)\ket{\bk}
\over
E - (\varepsilon_{\bK/2+\bq} + \varepsilon_{\bK/2-\bq})} \nonumber \\
&&
-\sum_{\bq}{\bra{\bk'}T_{\rm 2b}(\bar{E})\ket{\bq}
\bra{\bq}T_{\rm MB}(E)\ket{\bk}
\over
\bar{E} - (\varepsilon^{sp}_{\bK/2+\bq} + \varepsilon^{sp}_{\bK/2-\bq})}.
\label{eq:Tcorr}
\end{eqnarray}
If we now assume that there is a value of $\bar{E}=\bar{E}^*$ for which
$\bra{\bk'}T_{\rm MB}(E)\ket{\bk} = \bra{\bk'}T_{\rm 2b}(\bar{E})\ket{\bk}$, we
can replace $T_{\rm MB}(E)$ on the RHS of Eq.~(\ref{eq:Tcorr}) by $T_{\rm
2b}(\bar{E}^*)$.  The value of $\bar{E}^*$ may then be found by solving for
$\bra{\bk'}T_{\rm corr}(E,\bar{E}^*)\ket{\bk} = 0$.  We again take the limit of
zero temperature, such that $n_{\bK/2+\bq},n_{\bK/2-\bq}$ are zero, and for
collisions between two atoms in the condensate we take the limit $\bk,{\bf k'},
\bd{K},E =0$.  The value of $\bar{E}^*$ is then given in $D$ dimensions by the
solution to
\begin{equation}
0 = \int_0^\infty {k^{(D-1)} \over -2\varepsilon_k}dk -
\int_0^\infty {k^{(D-1)} \over \bar{E}^* - \hbar^2k^2/m} dk. \label{eq:Estar}
\end{equation}
Substituting the Bogoliubov dispersion relationship for the quasiparticle
energies using Eq.~(\ref{eq:bogenergies}) and carrying out the integrals
in Eq.~(\ref{eq:Estar}) we can obtain expressions for $\bar{E}^*$.  We are
then able to express the coupling parameter which occurs in the GPE in terms of
the two-body T-matrix evaluated at the energies $\bar{E}^*$.  In two and three
dimensions this leads to
\begin{equation}
g = \bra{\bd{0}}T_{\rm MB}(0)\ket{\bd{0}} = \left\{
\begin{array}{ll}
\bra{\bd{0}}T_{\rm 2b}(-\mu)\ket{\bd{0}}& \quad \mbox{in 2D,} \\
\bra{\bd{0}}T_{\rm 2b}(-{16\over \pi^2}\mu)\ket{\bd{0}}& \quad \mbox{in 3D.}
\end{array} \right. \label{eq:Estarvalues}
\end{equation}
However, in 1D the situation is more complicated because the first integral is
logarithmically divergent.  This case will be dealt with in
section~\ref{sec:1D} where we show that the results obtained are consistent
with known exact results. Prior to that however, we derive in the following
section the form of the many-body T-matrix in two dimensions and show that the
effective interaction energy becomes important in this case.

\subsection{The many-body T-matrix in two dimensions}

\label{sec:2dTmatrix}

We consider the case of a 2D Bose gas with an interatomic potential $V(\brho)$
which is short-range, parameterised by a length $a_{\rm 2d}$, and which admits
no bound states.  Specifically, we consider the case of a `hard-disc' potential
such that $V(\brho) = \infty$ for $|\brho| \leq a_{\rm 2d}$ and $V(\brho) =0$
otherwise. In recent work~\cite{scattering} we have derived a full expression
for the two-body T-matrix for this potential in the general off shell case.  In
the limit $ka_{\rm 2d}, k'a_{\rm 2d} \ll 1$ the result is
\begin{equation}
\bra{\bk'}T_{2b}(E)\ket{\bk} = 
{4\pi\hbar^2/m \over \pi i -2\gamma_{\rm EM} - \ln(Ema_{\rm 2d}^2/8\hbar^2) }, 
\label{2dsam}
\end{equation}
where the corrections are of order $(ka_{\rm 2d})^2$ and $1/\left[\ln(ka_{\rm
2d})\right]^3$ or greater, and $\gamma_{\rm EM}$ is the Euler-Mascheroni
constant. This result agrees with the work of Stoof~\cite{stoof1988}, and also
in the half-on-shell limit with  the earlier work of Schick~\cite{schick} and
Bloom~\cite{bloom}.  It is also of the same form as the results obtained by
Fisher and Hohenberg~\cite{fisher} who considered the case of a Gaussian
interatomic  potential, implying that the result is general for most
short-range repulsive potentials which may be parameterised by a length $a_{\rm
2d}$.  In this low momentum limit the T-matrix is independent of both $\bk$ and
${\bf k'}$ and thus it is still represented in position space by a delta
function effective interaction potential.  The new feature compared to the 3D
case is that the T-matrix now depends on energy and in particular it vanishes
as $E \rightarrow 0$.  It is therefore crucial to take into account the
many-body shift in the effective collision energy of two condensate atoms. 
This is now a self-consistent problem as many-body effects give rise to a
non-zero coupling constant.  In 3D the two-body T-matrix is non-zero as $E
\rightarrow 0$ and many-body effects can therefore be neglected at leading
order for dilute gases.

From equations~(\ref{eq:Estarvalues}) and~(\ref{2dsam}) the many-body T-matrix,
and therefore the coupling parameter, in 2D is found to be
\begin{equation}
g_{\rm 2d} = \bra{{\bf 0}}T_{\rm 2d}(-\mu)\ket{{\bf 0}} = 
-{4\pi \hbar^2 \over m}{1 \over \ln\left(\mu ma_{\rm 2d}^2/4\hbar^2\right)},
\label{eq:TMBapprox}
\end{equation}
where terms of order $1/[\ln(\mu ma_{\rm 2d}^2/4\hbar^2)]^2$ or greater have
been neglected.  Note that the evaluation of the two-body T-matrix at a
negative energy means that the imaginary component in Eq.~(\ref{2dsam})
vanishes, and thus the many-body T-matrix is real.

The parameter which appears in this description of the interparticle
interactions is the two dimensional scattering length $a_{\rm 2d}$, analogous
to the 3D \emph{s}-wave scattering length $a_{\rm 3d}$ which parameterises
three dimensional collisions in cold dilute gases.  Reliable values of $a_{\rm
3d}$ have been obtained in 3D by experimental measurements, and potentially
$a_{\rm 2d}$ could be measured in this manner. However, in their work on
quasi-2D scattering processes Petrov and Shlyapnikov~\cite{petrov2001} also
derived an expression for this parameter in terms of the three dimensional
$a_{\rm 3d}$, and the confinement of the trap in the tight direction $l_z$ 
\begin{equation} 
a_{\rm
2d} = 4\sqrt{\pi \over B}l_z \exp\left(-\sqrt{\pi}{l_z \over a_{\rm 3d}}\right),
\label{eq:a2d}
\end{equation} 
where $B \approx 0.915$. Using this expression for $a_{\rm 2d}$ in
Eq.~(\ref{eq:TMBapprox}) we obtain the quasi-2D coupling parameter of
Petrov and Shlyapnikov given in Eq.~(\ref{eq:gq2d}), and our approach
therefore agrees with their results in the genuine 2D limit which is
appropriate for $l_z \lesssim a_{\rm 3d}$.

Using this expression we are able to compare the strength of the 2D and
quasi-2D coupling parameters with  the parameter for quasi-2D gases with 3D
scattering $g'_{\rm 3d}$ described in section~\ref{sec:quasi2d} .  These
quantities are displayed as a function of trap width in the $z$ dimension in
Fig.~\ref{fig:quasi2dgs}.  It can easily be seen that the size of coupling
parameter appearing in the GPE for the genuine 2D case is over an order of
magnitude greater than in the case where the scattering is essentially 3D in
nature ($l_z/a_{\rm 3d} \gg 1$).  The magnitude of $g_{\rm 2d}$ decreases
slowly as $l_z$ is decreased beyond $\sim a_{\rm 3d}/2$ (not shown on the
graph) due to the size of $a_{\rm 2d}$ determined from Eq.~(\ref{eq:a2d}),
and it matches the 3D scattering limit for $l_z/a_{\rm 3d} \gtrsim 10$.

\begin{figure}
\epsfig{file=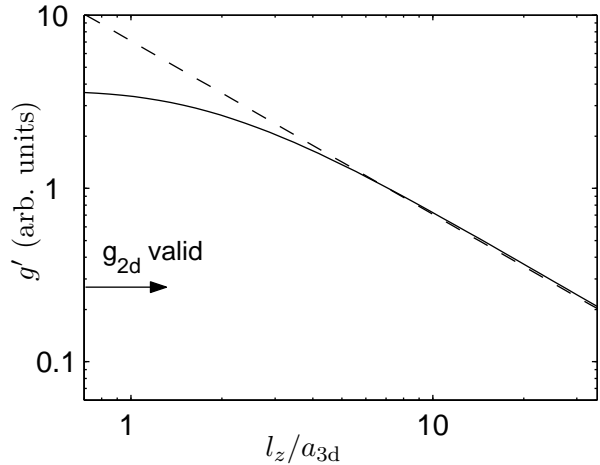,width=8cm}
\caption{Log-log graph of the effective 2D interparticle interaction strength
as a function of confinement in the third dimension. 
The solid line shows $g'_{\rm q2d}$ which
describes scattering in quasi-2D gases, taken from
ref.~\protect\cite{petrov2001}.  Our results for $g_{\rm 2d}$ derived in this
paper are consistent with this result and were derived for the region of
validity shown.  The dashed line shows $g'_{\rm 3d}$ which is the expected limit
at large $l_z/a_{\rm 3d}$.\label{fig:quasi2dgs}}
\end{figure}

\section{The many-body T-matrix in one dimension}

\label{sec:1D}

Before we use the result in the previous section to solve the two dimensional
GPE, we briefly consider the one dimensional case.  Our discussion in this
section is not intended to be rigorous, but is meant instead to demonstrate the
importance of including many-body effects, via the many-body T-matrix, when
considering the properties of a Bose gas in low dimensions.

A one dimensional condensate is described by the Gross-Pitaevskii equation
\begin{equation}
-{\hbar^2 \over 2m} \nabla^2 \psi(x) + V_{\rm trap}(x) \psi(x) +
N_0 
g_{\rm 1d}\left|
\psi(x)\right|^2 \psi(x) = \mu \psi(x), \label{eq:1DGPE}
\end{equation}
where $g_{\rm 1d}$ is the one dimensional coupling parameter.  The use of the
GPE necessarily assumes the existence of a condensate, which in one dimension
implies that the system must be confined in a trap and therefore of a finite
size. In a homogeneous 1D system a true condensate may not exist in the
thermodynamic limit due to the density of
states~\cite{hohenberg1967,lenard1966}.  With this caveat in mind we will use
the 1D case to illustrate the importance of the energy dependence of the
many-body T-matrix.  Specifically we will consider the one dimensional analogue
of a hard sphere gas for which exact results exist. This gas has an interatomic
potential of the form
\begin{equation}
V(x) = \left\{\begin{array}{ll}
0 &\quad \mbox{for } |x| >a_{\rm 1d} \\
\infty&\quad \mbox{for } |x| \leq a_{\rm 1d}.
\end{array}\right.
\end{equation}

In a recent paper~\cite{scattering} we have used an inhomogeneous
Schr\"{o}dinger equation to obtain results for the general off-shell two-body
T-matrix for hard sphere gases in one, two, and three dimensions.  In one
dimension in the limit of zero momenta the result is
\begin{equation}
\bra{{\bf 0}}T_{\rm 2b}(E)\ket{{\bf 0}} = \left\{
\begin{array}{ll}
-{2\over a_{\rm 1d}}\left( {\hbar \over \sqrt{m}}i
\sqrt{E}a_{\rm 1d} + Ea_{\rm 1d}^2 \right) &
\mbox{for } E>0, \\
{2\over a_{\rm 1d}}\left( {\hbar \over \sqrt{m}}
\sqrt{|E|}a_{\rm 1d} - Ea_{\rm 1d}^2 \right) &
\mbox{for } E<0. \\
\end{array} \right.
\end{equation}
As in the two dimensional case, the T-matrix is dependent on the collision
energy even at lowest order, and so the shift to an effective interaction
energy predicted in section~\ref{sec:negenergy} due to many-body effects is
again important.  Furthermore, in the case that $Ea_{\rm 1d}^2 \ll 1$, the
leading order term in the T-matrix in 1D is imaginary if $E$ is positive.  The
shift to a \emph{negative} effective interaction energy is therefore critical
in this one dimensional case.

In order to obtain the many-body T-matrix in terms of this two-body T-matrix we
must solve Eq.~(\ref{eq:Estar}). As noted earlier the first integral in
this equation is logarithmically divergent in 1D.  Physically this arises from
the fact that a true condensate does not exist in a homogeneous 1D system. 
Instead of a single quantum level with a macroscopic occupation (as occurs in a
true condensate), in 1D there is a band of low energy levels which all have
large occupations.  The same methods as discussed above may still be used in
the 1D case however, provided that we now define the ``condensate'' as a band
of levels in momentum space up to a cut-off at $k_{\rm max}$, such that the 1D
``condensate density'' $n_0 \equiv \sum_{k_i<k_{\rm max}} n_i$ satisfies $n_0
\sim n$ (as for a true condensate).  Using this definition, the lower limit of
the first integral in Eq.~(\ref{eq:Estar}) should then be $k_{\rm max}$
and the divergence is removed.  This approach is justified for a confined 1D
system since we may assume the existence of a condensate due to the
modification to the density of states which also removes the divergence.

A reasonable value for $k_{\rm max}$ may be obtained from the momentum
distribution for a system of impenetrable bosons.  Such a distribution is
discussed in Ref.~\cite{olshanii1998}.  We will define $k_{\rm max}$ by the
criterion that $N(k> k_{\rm max}) < 1$, which gives $k_{\rm max} \approx
0.25\pi n_0$ \cite{olshanii1998}.  Using this as the cut-off in
Eq.~(\ref{eq:Estar}), a solution may be found for $\bar{E}^*$ by making
the ansatz that $\bar{E}^* =-C\mu$, where $C$ is a constant.  For
the hard sphere case considered here the ansatz is satisfied when $C$ is the
solution to
\begin{equation}
\tanh^{-1}\left[{2 \over
\sqrt{(0.25\pi)^2/4C + 4}}\right] = { \pi^2\over C}. 
\end{equation}
This can be solved numerically to give $C\approx 3.4$. The expression for the
many-body T-matrix in 1D to leading order is then 
\begin{equation}
g_{\rm 1d} = \bra{{\bf 0}}T_{\rm
2b}(-C\mu)\ket{{\bf 0}} = \sqrt{4C\hbar^2 \mu/m}. \label{eq:TMB1D}
\end{equation}

We now consider a homogeneous 1D Bose gas, using the above definition of the
condensate.  When this system is in the ground state the contributions in the
GPE from the curvature of the wave function and the trapping potential both
vanish, and therefore
\begin{equation}
\mu_{\rm 1d} = n_0g_{\rm 1d} = {4C \hbar^2 \over m}n_0^2.
\end{equation}
This form differs from that found in 3D where $\mu \propto n_0$ because of the
dependence of $g_{\rm 1d}$ on the chemical potential.

This result can also be explained heuristically, as the extra curvature
introduced into the wave function by the presence of the other atoms.  If we
consider a many-body wave function which scatters off a hard sphere potential
of range $a_{\rm 1d}$, then in the limit of zero energy, we need to solve $d^2
\psi /dx^2 =0$.  We impose the boundary conditions that $\psi(x) =0$ at
$x=a_{\rm 1d}$ and $\psi(x)$ approaches an asymptotic value $\chi$ at large
$x$.  Since $\psi$ is a many particle wave function, the distance at which it
must arrive at its asymptotic value will be of the order of the interparticle
spacing $l_0$.  This gives a solution to the scattering problem of
\begin{equation}
\psi(x) = {\chi \over l_0 -a_{\rm 1d}} (x-a_{\rm 1d}) 
\qquad \mbox{for } a_{\rm 1d}<x\lesssim l_0.
\end{equation}
The extra energy caused by the curvature of the wave function
in this region is then
\begin{equation}
-{\hbar^2 \over 2m}\int_a^{l_0} |\nabla\psi(x)|^2 dx \approx -{\hbar^2|\chi|^2
\over 2ml_0}.
\end{equation}
And since $l_0 = 1/n_0$ and $|\chi|^2=n_0$ we have that the interparticle
interactions make a contribution to the energy which scales as $n_0^2$.  The
same result may be derived from an even simpler argument which considers each
particle to be confined in an infinite square well of length $\sim 1/n$ by its
nearest neighbours. 

The exact result for $\mu$ in such a 1D gas has long been
known.  In solving the system of 1D interacting bosons by demonstrating
equivalence with a gas of 1D non-interacting fermions,
Girardeau~\cite{girardeau1960} showed that in the strong coupling limit
(appropriate to the hard-sphere potential considered above)
\begin{equation}
\mu = {\pi^2 \over 2} {\hbar^2 \over m} n^2.
\end{equation}

Our result therefore shows the correct dependence on $n^2$, but disagrees on
the numerical factor.  The disagreement is due to the fact that, as previously
mentioned, in a homogeneous 1D system there can never be a true Bose
condensate, so significant corrections to the GPE can be expected.  The
additional uncertainty in the choice of $k_{\rm max}$ also introduces a source
for discrepancy in the numerical factor.  However, the agreement with the
dependence on $n^2$ indicates that the energy dependent many-body T-matrix
appears to deal with the interactions correctly.  This is interesting because
it means that an intrinsically many-body effect, namely particle confinement by
neighbours, can be modelled by an off-shell two-body T-matrix evaluated at a
shifted effective interaction energy which is the essential argument of this
paper.  This suggests that the method will have at least qualitatively the
correct density dependence in the strong coupling limit.  A more detailed
investigation of this approach in the 1D case will be the subject of a further
paper.  Although our discussion in this section has been qualitative due to the
lack of a true condensate in a 1D homogeneous system even at zero temperature,
in a trapped 1D system it is possible for a true condensate to form.  We
therefore expect that semi-quantitative results outside the normal BEC regime
of validity can be achieved using this method. In two dimensions a true
condensate can be formed, even in a homogeneous system, at $T=0$ and so we
expect our 2D results in this paper to be quantitatively correct.

\section{Scattering in Inhomogeneous Gases}

In the previous two sections we presented expressions for the many-body
T-matrix in one and two dimensions in terms of the two-body T-matrix evaluated
at shifted effective interaction energies.  However, the results obtained are
strictly only valid for homogeneous systems since we have not accounted for any
modifications of the scattering wave functions due to the presence of a
confining potential.  We consider here the case of a gas confined tightly
in one or two dimensions (in order to reduce the dimensionality, as discussed
in section~\ref{sec:quasi2d}) and weakly in the remaining dimensions on a
length scale $l_{\rm trap}$.

Provided that the range of the interatomic potential $R_e$ is much smaller than
$l_{\rm trap}$ then the scattering will be locally homogeneous and we can
replace $\mu$ where it occurs in equations~(\ref{eq:TMBapprox})
and~(\ref{eq:TMB1D}) by the homogeneous expression $\mu = n_0g$.  This is a
form of local density approximation and, as the density of an inhomogeneous gas
is spatially dependent, this leads to spatially dependent coupling parameters.
Recognising that $n_0(\br) = N_0|\psi(\br)|^2$ the coupling parameters in one
and two dimensions are
\begin{eqnarray}
g_{\rm 1d} &=& {4 C \hbar^2 N_0 \over m} |\psi(x)|^2 \label{eq:g1dtrapped}\\
g_{\rm 2d} &=& -{4\pi\hbar^2 \over m} \left[ \ln(N_0\pi |\psi(\brho)|^2a_{\rm
2d}^2)
\right]^{-1} +o\left(\ln[\ln(n_0a_{\rm 2d}^2)]\over \ln(n_0a_{\rm 2d}^2)\right) 
\label{eq:g2dtrapped}
\end{eqnarray}
These results agree with the work of Kolomeisky \emph{et al.\
}~\cite{kolomeisky2000,kolomeisky1992} who obtained similar expressions based
on a renormalisation group analysis.  Such density dependent coupling
parameters are also expected from the results of density functional
theory~\cite{nunes1999} which predict that the energy of the system is a
functional of the density only.  The same results may be obtained from
mean-field theory by incorporating the spatially dependent anomalous average
$\langle \hat{a}_i\hat{a}_j \rangle$ into the system of equations governing a
condensate and solving self-consistently~\cite{GHFB}.

\section{2D Solutions of the Non-linear Schr\"{o}dinger Equation}
\label{sec:2Dnumerical}

In this section we present solutions of the GPE for a trapped two dimensional
gas.  The solutions are found for a given $\mu$ by propagating the time
dependent GPE forward in imaginary time from an initial approximate solution to
obtain both the ground state wave function and the non-linearity $g_{\rm
2d}N_0$.  As mentioned in the previous section the coupling parameter in two
dimensions in a trap is spatially dependent, having a logarithmic dependency on
the density.  However, since in two dimensions the spatial dependence is merely
logarithmic it will have little effect on the solutions of the GPE, except at
the very edges of the trap where the wave function vanishes.  We therefore use
the homogeneous system coupling parameter of Eq.~(\ref{eq:TMBapprox}),
which will illustrate the features of most interest.

Using the expression for the 2D coupling parameter found in
Eq.~(\ref{eq:TMBapprox}) we solve the two dimensional time-independent
Gross-Pitaevskii equation for a 2D Bose condensate in a trap with $V_{\rm
trap}(\brho) = {1 \over 2} m\omega^2\brho^2$.  
We can make the GPE dimensionless, scaling all energies by $\hbar\omega$
and all lengths by $l_{\rho} = \sqrt{\hbar/(2m\omega)}$, giving
\begin{equation} -\tilde{\nabla}^2\psi(\tilde{\brho}) + \tilde{V}_{\rm
trap}(\tilde{\brho})\psi(\tilde{\brho}) + N_0 
\tilde{g}_{\rm 2d}(\tilde{\mu})\left|
\psi(\tilde{\brho})\right|^2\psi(\tilde{\brho}) = \tilde{\mu}
\psi(\tilde{\brho})
\label{eq:dimlessGPE},
\end{equation}
where $\tilde{g}_{\rm 2d}(\tilde{\mu}) = -8\pi/\ln(\tilde{\mu}\tilde{a}_{\rm
2d}^2/8)$ and $\tilde{V}_{\rm trap}(\tilde{\brho}) = {1\over
4}\tilde{\brho}^2$. Note that the quantity $\tilde{\mu}\tilde{a}_{\rm 2d}^2$ is
small compared to unity (or the earlier expansion of the T-matrix elements
fails) and therefore the interaction is repulsive.  
As shown earlier, for
the range of $\mu$ and $N_0$ that we consider here, we find that
Eq.~(\ref{eq:TMBapprox}) leads to a value for $\tilde{g}_{\rm
2d}(\tilde{\mu})$ which is more than an order of magnitude greater than the
equivalent value for a quasi-2D gas in which the particle interactions are
effectively 3D in nature.  Thus the non-linear term in the GPE is more
significant in 2 dimensions than in the 3D case.

\subsection{Ground state solutions}

Figure~\ref{fig:samplesolutions} presents sample solutions for the ground state
of a 2D BEC in a trap for differing values of $\tilde{\mu}$.  To illustrate the
physical quantities involved we give numbers for a gas in a trap of $\omega =
2\pi \times 100{\rm Hz}$ and with a scattering parameter given by $a_{\rm 2d} =
6\,{\rm nm}$.  This is close to the 3D \emph{s}-wave scattering length $a_{\rm
3d}$ found for $^{87}{\rm Rb}$, and therefore from Eq.~(\ref{eq:a2d}) this
corresponds to a situation where $l_z \approx a_{\rm 3d}$.  We see that at low
$N_0$ the solution is approximately the Gaussian wave function that is 
expected for the non-interacting case.  At higher $N_0$  the Thomas-Fermi
approximation  found by neglecting the contribution to the GPE from the kinetic
energy term as compared to the interaction and trapping terms is expected to be
a good description.  In two dimensions the Thomas-Fermi approximation gives a
density profile in the form of an inverted parabola

\begin{equation}
\left|\psi(\tilde{\brho})_{TF}\right|^2 = 
-{\ln(\tilde{\mu}\tilde{a}_{\rm 2d}^2/8) 
\over N_0
8 \pi}\left[\tilde{\mu} - \tilde{V}_{\rm trap}(\tilde{\brho}) \right] 
\theta\left(\tilde{\mu}-\tilde{V}_{\rm trap}(\tilde{\brho})\right),
\end{equation}
where $\theta(x)$ is the step-function.  At higher $N_0$ the solutions shown
are generally very well approximated by the Thomas-Fermi form, except at the
boundary region of the condensate.  Indeed we find that the Thomas-Fermi
approximation works well for the 2D case, due to the high strength of the
scattering, as expected from the dimensionless GPE~(\ref{eq:dimlessGPE}). For
the non-linear term to dominate the kinetic energy term requires that $N_0^{2d}
\gg -\ln(\tilde{\mu}\tilde{a}_{\rm 2d}^2/8)$ (where $\tilde{\mu} \sim 10-100$),
whilst in the three dimensional case we require $N_0^{3d} \gg 1/\tilde{a}_{\rm
3d}$.  Putting typical numbers into this using our parameters we get $N_0^{2d}
\gg 10$, whilst $N_0^{3d} \gg 100$, and thus the Thomas-Fermi regime is reached
in 2D with about an order of magnitude fewer atoms than is the case for 3D.  As
confirmation of this, the Thomas-Fermi approximation for the number of
condensate atoms is $N_0 = -\mu^2\ln\left(\mu m a_{\rm
2d}^2/4\hbar^2\right)/(2\hbar\omega)^2$ and is found to be in agreement with
the numerical results to within one percent once $N_0$ was greater than $300$.

\begin{figure}
\epsfig{file=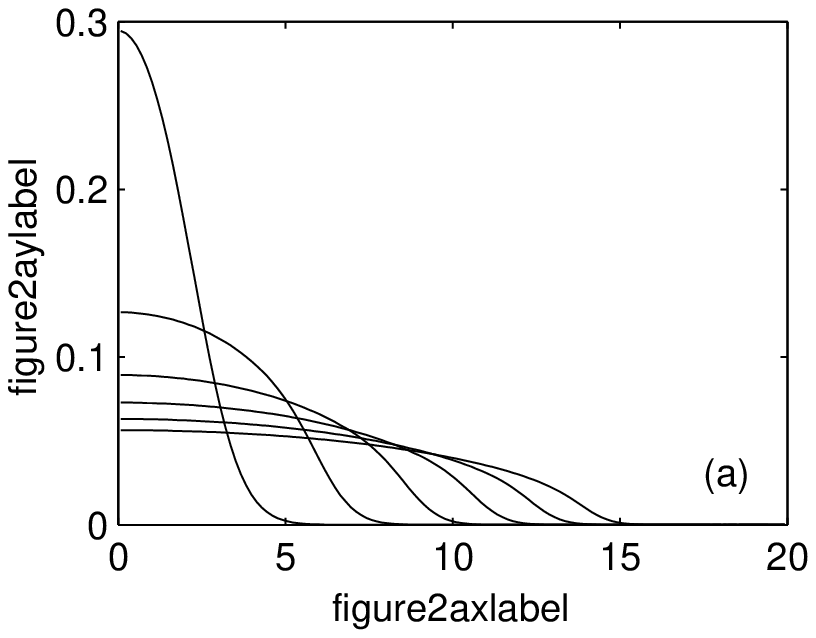,width=8cm}
\epsfig{file=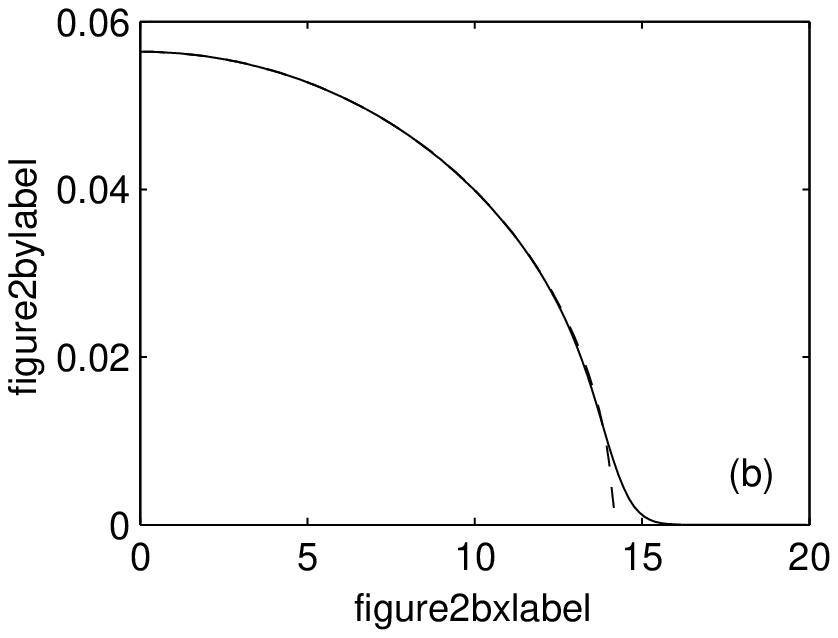,width=8cm}
\caption{(a) Ground state  2D GPE solutions in an axisymmetric trap for $\mu =
2,10,20,30,40$, and $50 \hbar\omega$.  $\psi(\tilde{\brho})$  is normalised to
unity, and populations given assume parameters $\omega = 2\pi \times 100{\rm
Hz}$, $a_{\rm 2d} = 6\,{\rm nm}$. (b) Comparison of GPE solution for
$\tilde{\mu}=50$ (solid line) with Thomas-Fermi approximation (dashed line).
\label{fig:samplesolutions}}
\end{figure}

In some previous papers~\cite{kim} the GPE has been solved with $g_{\rm 2d}$
approximated by  an energy independent constant.  This is appropriate to the
case where the scattering is three dimensional, but not to the fully 2D case
where $g_{\rm 2d}$ depends on $\mu$.  We find here that the interaction
strength given by Eq.~(\ref{eq:TMBapprox}) increases by about 50\% as
$\mu$ rises from $2\hbar\omega$ to $50\hbar\omega$.  Figure~\ref{fig:muindep}
shows the possible errors which can arise from making the assumption of a
constant coupling parameter.  Each line plotted on this graph assumes a
constant $g_{\rm 2d}$, the strength of which is chosen to agree with
Eq.~(\ref{eq:TMBapprox}) at a certain value of the chemical potential
$\mu_*$.  The figure shows that results obtained with a constant $g_{\rm
2d}(\mu_*)$ will introduce systematic errors when $\mu$ is significantly
different from $\mu_*$.  In the Thomas-Fermi approximation the relative error
incurred in a measurement of $N_0$ assuming a constant $g_{\rm 2d}(\mu_*)$ is
given by $\ln(\mu_*/\mu)/\ln(\tilde{\mu} \tilde{a}_{\rm 2d}^2/8)$.

\begin{figure}
\epsfig{file=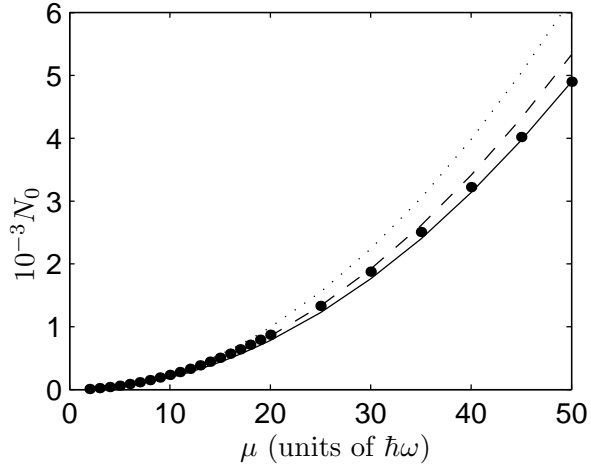,width=8cm}
\caption{$N_0$ vs $\mu$ for a 2D Bose gas.  The dots represent solutions of
the GPE with the full energy dependent interaction given in
Eq.~(\protect\ref{eq:TMBapprox}).  The lines are results which assume
a constant (independent of $\mu$) coupling parameter
$g_{\rm 2d}$.  
The three constant values of $g_{\rm 2d}$ correspond to 
Eq.~(\ref{eq:TMBapprox}) evaluated at $\mu$ equal to $6\hbar\omega$
(dotted),
$25\hbar\omega$ (dashed) and  $50\hbar\omega$ (solid).\label{fig:muindep}}
\end{figure}

\subsection{Vortex state solutions}

The 2D GPE can also be solved for the case of a two-dimensional condensate in a
symmetric trap containing a vortex at the centre by
looking for solutions of the form
\begin{equation}
\psi(\brho) = \phi(|\rho|)e^{i\kappa\theta},
\end{equation}
where $\theta$ is the angle around the vortex core, and the phase wraps around
by $2\pi\kappa$, where $\kappa$ is an integer, as the range of $\theta$ is
traversed. This adds an ``effective potential'' to the GPE and we now solve
\begin{equation}
-{\hbar^2 \over 2m} \nabla^2 \phi(\rho) + {\hbar^2 \kappa^2 \over 2mr^2}+ 
V_{\rm trap}(\rho) \phi(\rho) + N_0 g_{\rm 2d}(\mu) \left|
\phi(r)\right|^2 \phi(\rho) = \mu \phi(\rho).
\end{equation}
Solutions of these vortex states are shown in Fig.~\ref{fig:vortexprofiles}.

\begin{figure}
\epsfig{file=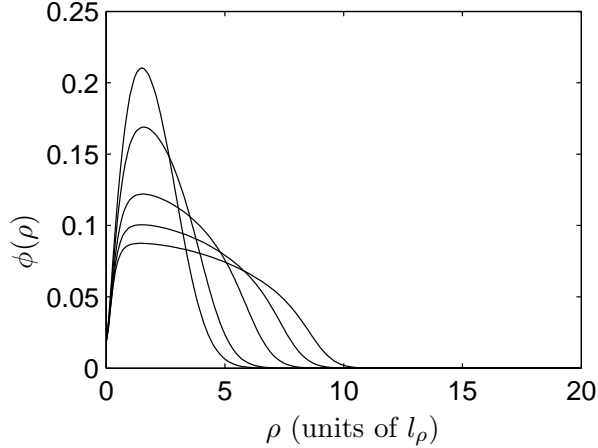,width=8cm}
\caption{Sample 2D GPE solutions for a vortex state with $\kappa =1$ and for
values of $\mu$ of $3$, $5$, $10$, $15$, and $20\hbar\omega$.  
$\phi(\rho)$ is
normalised to unity, and populations given assume parameters $\omega = 2\pi
\times 100 {\rm Hz}$, $a_{\rm 2d} = 6{\rm nm}$. \label{fig:vortexprofiles}}
\end{figure}

Such vortex states, which carry an angular momentum $L_z = N\hbar\kappa$, can
be made energetically favourable by rotating the trap with sufficiently high
frequency $\Omega$.  The energy functional for a wave function in the
non-rotating frame is 
\begin{equation}
E[\psi] = \int d\brho \left[ {\hbar^2 \over 2m}|\nabla \psi(\brho) |^2 + V_{\rm
trap}(\brho)|\psi(\brho)|^2 + {g_{\rm 2d}(\mu) \over 2}|\psi(\brho)|^4
\right].
\end{equation}

The point at which $E[\psi_{\kappa =1}] -\Omega L_z$ becomes less than
$E[\psi_{\kappa =0}]$ is known as the thermodynamic critical frequency, and
this is plotted in Fig.~\ref{fig:criticalfreq} for both 2D and (genuine) 3D
condensates.  The three dimensional results were calculated from solutions of
the 3D GPE, given by Eq.~(\ref{eq:3DGPE}), with $a_{\rm 3d}$ taken to be
equal to $a_{\rm 2d}$ the scattering length used for the 2D results.
Creation of a vortex in the centre of the trap comes at the cost
of increasing the contributions from both the kinetic energy and the trapping
potential terms in the GPE, although the non-linear contribution is reduced by
virtue of a lower central density.  Stronger nonlinear systems are therefore
more susceptible to vortex creation, and this becomes energetically favourable
at much lower frequencies in 2D than in 3D for the same value of the scattering
length $a$, as seen in Fig.~\ref{fig:criticalfreq}.

\begin{figure}
\epsfig{file=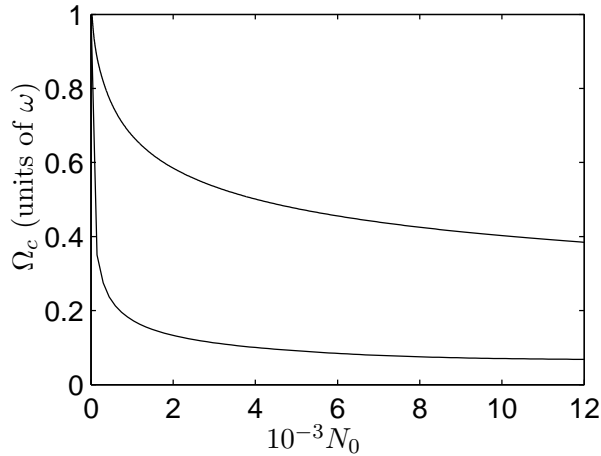,width=8cm}
\caption{The critical frequency $\Omega_c$ at which vortex formation becomes
energetically favourable in 2D (lower) and 3D (upper) gases as a function of
condensate population $N_0$.  Results obtained using $a_{\rm 3d} = a_{\rm 2d}$.
\label{fig:criticalfreq}}
\end{figure}

\section{Discussion and Conclusions}

In this paper we have found expressions for the many-body T-matrix in a dilute
Bose gas describing the collisions occurring in a condensate in terms of the
simpler two-body T-matrix.   We have shown that many-body effects of the
condensate mean-field on such collisions may be incorporated by a shift in the
effective interaction energy of a two-body collision, and that such an approach
leads to the same results obtained from renormalisation group
techniques~\cite{kolomeisky2000,kolomeisky1992}.

The fundamental difference to the three dimensional
case is that the first order term in the T-matrix in lower
dimensions is dependent not only on the scattering length, but
also on the energies of the colliding particles. The coupling parameter in one
and two dimensions is therefore dependent on the chemical potential of the
condensate.

The energy dependent form of the many-body T-matrix in 2D found here can be
used to obtain a self-consistent form for the 2D Gross-Pitaevskii equation.  We
have presented sample solutions and have shown that the importance of the
nonlinear term is magnified in 2D (as compared to the 3D case) due to the size
of the coupling constant in two dimensions.  The Thomas-Fermi approximation is 
therefore valid at a much lower number of atoms than in the 3D case,
approximately an order of magnitude lower in the case considered here. The
critical frequency of vortex formation is also found to decrease with
condensate occupation much faster in 2D than in 3D, and so vortices should be
comparatively easier to form in 2D.

\acknowledgments  
This research was supported by the Engineering and Physical Sciences Research
Council of the United Kingdom, and by the European Union via the ``Cold Quantum
Gases'' network.  In addition, M.D.L.\ was supported by the Long Studentship from
The Queen's College, Oxford, and S.A.M.\ would like to thank Trinity College,
Oxford, for financial support.

\end{document}